# On geometric delusions of hexagonal structures

M. J. I. Khan[*], S. Babar[*]

[*]Laboratory of computations and High Energy Physics
Bahauddin Zakariya University, Multan 60800, Punjab, Pakistan.
Email: drjunaid.iqbalkhan@bzu.edu.pk

## Abstract

The confining geometries of fermions in 2D structures exhibits interesting results that have highest symmetry. Delusion can be considered as the topological effect which is topological invariant. Topologically, genus zero surfaces needs excess of pentagons while in surfaces $g > 2$ surfaces needs excess of heptagons. The curvature effect and the rise of effective gauge field can be interpreted from delusion effects in hexagonal lattice. This idea is novel in its scope as it can state theoretical description of structures and their stability.

# Contents

## 1. Introduction

Two dimensional (2D) physics is very rich in condensed matter physics. In 2D systems, we can speculate the fermions dynamics by dirac equation [1]. The confining geometris of fermions in 2D structures exhibits interesting results that have highest symmetrical geometry. The most alluring pictures could be circle, triangle and hexagons where circle has the maximum symmetry and triangle has lowest [2]. The geometrical changes results when the lattice or system of fermions is subjected under the external fields. Under the action of these external fields, for example magnetic field, curvature is produced which could be negative or positive [3]. Geometric variations can produce topology [4] and from physical point of view topology helps to study gauge field and scalar fields [5]. The gauge field from strain in graphene has been studied by F. Juan et al. in [6]. The hamiltonian describing the low energy spectrum of these systems under fields can be described by dirac equation. Dirac operator can be constructed by studying the geometry of the 2D surface that might be appeared as the effect of polygonal deformations. Where polygon can be geometrically defined as patterns of pentagons and heptagons. By employing pentagons and heptagons, one can generate 2D hexagonal surface. The most promising example of 2D materials is grephene.

Graphene is a system of carbon atoms forming a hexagonal lattice delibrating a system in which carbon atoms are $sp^2$ hybridized. Graphene topology is a theoretical premises and its topology in 2D, 1D, 0D determines the electronic properties. It leads to interesting horizons in a wide range of study and its applications in electronics, spintronics, sensors, catalysts in chemistry and biology. The study of topological character of graphene provides foundations to address basic questions related to topological insulators, electronics properties and generalization of other 2D graphene like materials. Geometric deformations in 2D graphene lattice are introduced such that the heptagons and pentagons are adjusted leaving the 2D plane undisturbed. The wrapped 2D topology of regular hexagon is identified as the torus and has been discussed in [5].

The topological characteristics of the hexagonal 2D surfaces can be studied with the help of Euler theorem. It holds good for polyhadrons except the curved or compact surfaces which are closed, for example circle or ellipse. Euler characteristic is a topological invariant.

This paper is organized as follows. First section covers an introduction to the subject of study. In section 2, we gives realizations of the geometric variations in hexagonal structures. In section 3, we describe the phenomenon of delusion that causes a beautiful connection between geometry and topology. Finally, we end the paper in a way with a conclusion, in which we stated possible future work as a part of this study.

## 2. Geometric variations

Graphene is a plane consisting of a geometric pattern of regular hexagons. It is a carbon based material where carbon atoms are positioned at vertices of a two dimensional honeycomb lattice. In carbon based systems, 2D hexagonal structures sheets forms the basis for other structures like nanotubes, buckyballs, graphite [7]. Graphene shows exceptional electronic properties due to tunneling of electrons from one site of lattice to nearest one site. It is because of the flexibility of carbon valence electorns and resulting dimensionality of its bonding structures.

The low energy description of graphene can be picturized by determining total number of plaquettes. The important thing is to introduce the deformations in the hexagonal lattice such that the internal structure of lattice consisting of pentagons and heptagons should remain unchanged as they do not change the topology of the surface due to cancel out effect. In this case, 2D lattice have geometry,

$$[N_1(n_1 - 2) \times 180°] + N_2[(n_2 - 2) \times 180°], \tag{1}$$

Where $(N_1, N_2)$ are the number of heptagonal and pentagonal cells while $(n_1, n_2) = (7,5)$ for heptagons and pentagons respectively. In case, if topology of 2D sheet is restored then,

$$2N_1[(n_1 - 2) \times 180°] + 2N_2[(n_2 - 2) \times 180°] + N_3 \times 720°, \tag{2}$$

$N_3$ are the number of hexagonal cells that are present at the edge making edge states and lattice structure topological invariant.

Most common geometric variants of graphene are fullerence [8] and nanotubes [9-11]. For carbon nanotubes, as they have zero two dimensional curvature so there is no need for the presence of pentagons. The presence of pentagons build up between hexagons induces curvatures in the structures of hexagons. The physics of this coupling is fertile in the sense that it describe the physical properties of graphene based on different geometrical changes.

### 3. Connections between geometry and topology

The beautiful and nice description between deformations and topology is described by the Euler theorem. It gives an insight to the topological and structural properties together. These geometric delusion in hexagonal structures are topological invariants as the plane is introduced with couples of hexagons and heptagons. The Euler characteristic has a beautiful connection with topology,

$$\chi(g) = V - E + F = 2 - 2g. \tag{3}$$

It is for the orientable surface in a plane but this formula does not hold good for the non-orientable surface. So,

$$\chi(g) = 2 - g, \tag{4}$$

Where $g$ is the genus of the surface. Physically this idea of promoting topology induces an effective gauge field to the system. The interplay of pentagons and heptagons with the hexagons creates a delusion in the hexagonal lattice [12]. That creates curvature and resulting a external effective field [13]. Here physical meaning of delusion could be interpreted as the survival of hexagonal sheet to be flat surface which is meaningfull in non-trivial topology. Delusion is ruled by the topological change.

Hexagonal geometry plays a vital code for geometric delusions in hexagonal structures. It line up the mergence of atoms of pentagons to hexagons and vice versa. This phemenon can physically explain the stability of 2D hexagonal structures. Due to the presence of pentagons among hexagons, the interaction energy is supposed to be minimum when two atoms sit on the vertices of regular pentagons [14]. Delusion enforces atoms of pentagons and hexagons to keep the symmetry of the structure invariant but symmetry can be broken due to the presence of holes in the surface. Mergence of two pentagons could be a possibility which results in the inducing curvature effect in hexagons. On the other hand, their seperation indicates the higher $\pi$ electron stability [15].

Mathematically, manifolds or surfaces could be in dimensions superior to 2D so their requirement for the lattice symmetry is fulilled with the imbalance of pentagons and heptagons. Topologically, genus zero surfaces needs excess of pentagons while in surfaces $g > 2$ surfaces needs excess of heptagons.

For a hexagonal 2D lattice, due to presence of pentagons other than hexagons, curvature effects [16] are present that introduces gavitational aspects in theory of hexagons. Let $n_5$ be the number of pentagons that are supposed to appear in hexagonal plane. Then applying euler theorem in equ. (3), to these pentagons and hexagon, we can induce curvature effect that would twist geometry to buckyballs. On this 2D surface, let number of faces be $n_5 + n_6$ while the number of vertices be $V = (5n_5 + 6n_6)/3$. The number of edges is given by $E = (5n_5 + 6n_6)/2$ as edges are shared by two polygons. Substituting this in equ. (3) we obtain,

$$n_5 = 12(1 - g). \qquad (5)$$

This formula gives a nice intrepretation about the number of pentagons should be minimum 12 to keep the symmetry of the structure in a spherical form. In above equ. (5), if we have $g = 0$ the geometry correspond to fullerenes [17]. Pentagons and heptagons presence is excluded in genus one surfaces as they are equivalent to flat sheet. Geometric delusions can reproduce different geometric forms of graphene and in each geometry we find different physical properties. Topology induces a theoretical description of the hexagonal lattice that it may be in compact or non-compact form. For example, except fellerene the remaining geomertic structures of graphene are open surfaces.

**Conclusion**

Geometric delusions are phenomenon that leads to the introduction of a type of 2D lattice where the surface can be resistant to retain its physical properties in form of mergence of hexagons, pentagons and heptagons. Physically it notifies the significance of topology and an induced gauge field that is consistent with the theoretical description of graphene. The work is to be done to incorporate the physical introduction of induced or external gauge field in the theory that would couple to the gauge field which might be labelled as internal gauge field. There might be some additional external field background effects in theory.

**References**

1. Palumbo, G., R. Catenacci, and A. Marzuoli, *Topological effective field theories for Dirac fermions from index theorem.* International Journal of Modern Physics B, 2014. **28**(01).

2. Brack, M., et al., *On the role of classical orbits in mesoscopic electronic systems.* Zeitschrift für Physik D Atoms, Molecules and Clusters, 1997. **40**(1): p. 276-281.

3. Pachos, J.K., *Manifestations of topological effects in graphene.* Contemporary Physics, 2009. **50**(2): p. 375-389.

4. Cortijo, A., F. Guinea, and M.A. Vozmediano, *Geometrical and topological aspects of graphene and related materials.* Journal of Physics A: Mathematical and Theoretical, 2012. **45**(38): p. 383001.

5. Khan, M.J.I., Kamran, M., Babar, S., *On topological aspects of 2D graphene like materials.* Physica Scripta, 2014. **arXiv preprint**(arXiv:1408.6124v2).

6. de Juan, F., J.L. Mañes, and M.A. Vozmediano, *Gauge fields from strain in graphene.* Physical Review B, 2013. **87**(16): p. 165131.

7. Dodd Gray, A.M., Bhaskar Mookerji, *Crystal Structure of Graphite, Graphene and Silicon.* Physics for solid state applications, 2009. **6.730**.

8. Kroto, H.W., et al., *C 60: buckminsterfullerene.* Nature, 1985. **318**(6042): p. 162-163.

9. Ovidko, I., *Review on grain boundaries in graphene. Curved poly-and nanocrystalline graphene structures as new carbon allotropes.* Rev. Adv. Mater. Sci, 2012. **30**: p. 201-224.

10. Shenderova, O., V. Zhirnov, and D. Brenner, *Carbon nanostructures.* Critical Reviews in Solid State and Material Sciences, 2002. **27**(3-4): p. 227-356.

11. Radushkevich, L. and V. Lukyanovich, *About the structure of carbon formed by thermal decomposition of carbon monoxide on iron substrate.* J. Phys. Chem.(Moscow), 1952. **26**: p. 88-95.

12. Zsoldos, I., *Effect of topological defects on graphene geometry and stability.* Nanotechnology, science and applications, 2010. **3**: p. 101.

13. Pachos, J.K., A. Hatzinikitas, and M. Stone, *Zero modes of various graphene configurations from the index theorem.* The European Physical Journal-Special Topics, 2007. **148**(1): p. 127-132.

14. Vassilevich, D., *Heat kernel expansion: User's manual. Phys. Rept., 388: 279–360, 2003.* arXiv preprint hep-th/0306138. **27**: p. 25.

15. Sher, R. and R. Daverman, *Handbook of Geometric Topology*. 2001: Elsevier.

16. Pachos, J.K., M. Stone, and K. Temme, *Graphene with geometrically induced vorticity.* Physical review letters, 2008. **100**(15): p. 156806.

17. Itoh, S., et al., *Structure and energetics of giant fullerenes: An order-N molecular-dynamics study.* Physical Review B, 1996. **53**(4): p. 2132.